\documentclass[conference]{IEEEtran}
\IEEEoverridecommandlockouts
\usepackage{cite}
\usepackage{amsmath,amssymb,amsfonts}
\usepackage{multirow}
\usepackage{algorithm}
\usepackage{algpseudocode}

\usepackage{graphicx}
\usepackage{textcomp}
\usepackage{xcolor}
\def\BibTeX{{\rm B\kern-.05em{\sc i\kern-.025em b}\kern-.08em
    T\kern-.1667em\lower.7ex\hbox{E}\kern-.125emX}}
\begin{document}

\title{EthCluster: An Unsupervised Static Analysis Method for Ethereum Smart Contract\\
}

\author{\IEEEauthorblockN{1\textsuperscript{st} Hong-Sheng Huang}
\IEEEauthorblockA{\textit{Institute of Information Security} \\
\textit{National Tsing Hua University}\\
Hsinchu, Taiwan \\
ray.h.s.huang@gapp.nthu.edu.tw}
\and
\IEEEauthorblockN{2\textsuperscript{nd} Jen-Yi Ho}
\IEEEauthorblockA{\textit{Institute of Information Security} \\
\textit{National Tsing Hua University}\\
Hsinchu, Taiwan \\
vincent66@gapp.nthu.edu.tw}
\and
\IEEEauthorblockN{3\textsuperscript{rd} Hao-Wen Chen}
\IEEEauthorblockA{\textit{Institute of Information Security} \\
\textit{National Tsing Hua University}\\
Hsinchu, Taiwan \\
haowen@gapp.nthu.edu.tw}
\and
\IEEEauthorblockN{4\textsuperscript{th} Hung-Min Sun}
\IEEEauthorblockA{\textit{Department of Computer Science} \\
\textit{National Tsing Hua University}\\
Hsinchu, Taiwan \\
hmsun@cs.nthu.edu.tw}
}

\maketitle

\begin{abstract}
The emergence of virtual currencies and blockchain technology has introduced decentralized networks to the field of information privacy technology. Businesses and organizations have started to deploy their business logic services on the Ethereum blockchain, utilizing smart contracts that enable users to conduct transactions within the Ethereum network. These smart contracts often hold substantial amounts of virtual currency, drawing attention even from Wall Street due to the value of virtual currencies. This interest has made these contracts prime targets for hackers. Poorly designed smart contracts are particularly vulnerable, as they may allow at-tackers to steal the virtual currency managed by the contracts. Notably, once a smart contract is deployed, it cannot be modified, highlighting the importance of security inspections prior to deployment; Conducting a security assessment of smart contracts before their deployment is thus a widely adopted approach in program security. Current methods for inspecting general applications are typically categorized into static analysis, dynamic analysis, and machine learning detection. In Ethereum, all smart contracts are deployed on a decentralized network, which means that the source code of each contract is publicly accessible. Furthermore, it is possible to obtain both the Ethereum Virtual Machine bytecode and the Application Binary Interface (ABI) functions from the Ethereum public chain. In this work, we train a model using unsupervised learning to identify vulnerabilities in the Solidity source code for Ethereum smart contracts. Given the rarity of financial vulnerability incidents on Ethereum, we adopt an anomaly detection approach, defining our problem as a one-class classification task for each vulnerability type. Our model is de-signed to inspect real-world smart contracts from seven blockchain explorer platforms, including Etherscan, BscScan, and PolygonScan, which aggregate transaction records from the Ethereum public chain. To address the challenges posed by real-world smart contracts, our training data is based on actual vulnerability samples from datasets such as SmartBugs Curated and the SolidiFI Benchmark. These datasets enable us to construct a robust, unsupervised static analysis method to detect five specific vulnerabilities—Reentrancy, Access Control, Timestamp Dependency, tx.origin misuse, and Unchecked Low-Level Calls—using clustering algorithms to identify outliers, which are classified as vulnerable smart contracts.
\end{abstract}


\section{Introduction}
In 2008, after Satoshi Nakamoto introduced the first decentralized digital currency system, decentralized network architectures and blockchain technology rapidly developed and evolved. The concept of virtual currencies has now become deeply embedded in modern civilization. Blockchain frameworks such as Ethereum, Hyperledger, and EOS have been widely applied across various fields. Among these advancements, the emergence of smart contracts has enabled developers to write programs within blockchain environments. Businesses and organizations have started deploying their commercial logic on blockchains, allowing users to interact directly with the smart contracts deployed by enterprises. These smart contracts also store vast amounts of user data and virtual currencies.

However, due to the significant differences between the blockchain execution environment and traditional application environments, programmers may design vulnerable smart contracts. These vulnerabilities could allow malicious attackers to exploit them and steal virtual currencies stored within the smart contracts. The Ethereum network serves as a prime example, having experienced several major smart contract attacks since 2016, such as The DAO Attack, the Parity Wallet Attack, and Etherpot, among others. Notably, The DAO Attack resulted in the theft of 36 million Ether by attackers. To address this issue, Ethereum was forced to implement a hard fork, splitting the network into Ethereum Classic and Ethereum.

Considering this situation, NCC launched the Decentralized Application Security Project (DASP) \cite{b1} in 2018. This project identified the top 10 most discovered vulnerabilities in the Ethereum network, aiming to help developers of smart contracts incorporate security considerations into their development processes. Emphasizing security in this way will contribute to the continued development and application of blockchain technology. 

No matter the type of application being tested, static analysis remains a crucial and fundamental task. In this context, we examine two static analysis methods. The first method utilizes BERT \cite{b2}, fine-tuning a Natural Language Processing model to identify vulnerable sentences within smart contracts \cite{b3}. The second method employs a Graph Attention Network (GAT) \cite{b4}, training a neural network to predict whether a smart contract contains a reentrancy vulnerability.

In the NLP-based static analysis system, where the BERT natural language processing model is fine-tuned, the 50\% training set achieves an accuracy of 90\% in detecting smart contract vulnerabilities. Similarly, in the GAT-based static analysis system, which employs a Graph Attention Network (GAT) with an 80\% training and 20\% testing split, the self-attention mechanism also achieves an accuracy as high as 90\%. These results are expected, as both BERT and GAT are supervised models that rely on a large volume of labeled vulnerability samples to improve accuracy; Also in academic research, pre-labeled datasets of vulnerabilities are already available to support such models.

However, in the real world, actual attack incidents are rare, and as a result, malicious samples are also extremely scarce. This leads to an imbalance between the malicious data used for training and the large volume of test samples readily available in practical scenarios. To address this imbalance, unsupervised learning methods can be considered as an alternative. Rather than relying on extensive manual labeling, these methods identify patterns within the data during the training process to make determinations. This approach enables pseudo-labeling, allowing latent vulnerabilities in real-world smart contracts to be effectively uncovered. Therefore, we introduce this approach to detect vulnerabilities in Solidity smart contracts.

\section{Related Work}

\subsection{Un-supervised Learning}

Unsupervised learning techniques have demonstrated promising applications in software vulnerability detection \cite{b5}. Research by Ka-Wing Man \cite{b6} indicates that unsupervised learning methods can be effectively applied to predict software vulnerabilities. The study utilized Principal Component Analysis (PCA) and the k-means clustering algorithm to analyze metrics from software source code, enabling the prediction of vulnerabilities.

\subsection{Dimensionality Reduction}
In the study, Principal Component Analysis (PCA) \cite{b7} was applied for dimensionality reduction after feature extraction to reduce the data's dimensionality. PCA is a widely used dimensionality reduction technique that decreases the number of variables by extracting the principal components from the data, thereby retaining most of the information. The purpose of this step is to reduce data complexity and eliminate redundant information, making subsequent clustering analysis more accurate and efficient. The steps of PCA are as follows

\begin{enumerate}
    \item Calculate the (multivariate) covariance or correlation matrix from the sample data.
    \item Calculate the eigenvalues and eigenvectors of the covariance or correlation matrix.
    \item Generate principal components, where each principal component is a linear combination of the original variables with optimal weights. Here, $P_{i}$ is the $i$-th principal component, and $b_{i_{k}}$ is the regression coefficient of the variable $X_{k}$, as shown in the equation below: \begin{align}
        P_{i} = b_{i_{1}}X_{1} + b_{i_{2}}X_{2} + \cdots + b_{i_{k}}X_{k}
    \end{align}
\end{enumerate}

\subsection{Clustering}
The study uses the k-means clustering algorithm to perform clustering analysis on the dimensionally reduced data. The k-means algorithm is a commonly used unsupervised learning method \cite{b8}, which assigns data points into  clusters in such a way that data points within each cluster exhibit high similarity, while data points between different clusters show significant differences. In this study, k-means is used to identify different categories within the software source code, thereby distinguishing between vulnerable and non-vulnerable code. The steps of k-means are as follows:
\begin{enumerate}
    \item Select the number of $k$ clusters and partition the $n$ data points into $k$ clusters.
    \item Randomly initialize $k$ cluster centers by selecting points in the space as data points.
    \item For each data point, calculate the error to each cluster center and assign the data point to the cluster corresponding to the cluster center with the smallest error.
    \item For each cluster, calculate the mean of all the data points within that cluster. This mean serves as the new cluster center.
    \item Repeat steps 3 and 4 until the new cluster centers no longer change.
\end{enumerate}

\subsection{Feature Extraction}
Feature extraction refers to the process of extracting representative and distinguishing features from raw data, thereby simplifying the data processing and improving analysis efficiency. Feature extraction plays a critical role in smart contract vulnerability detection, as it effectively enhances the accuracy and efficiency of vulnerability identification.
    \begin{enumerate}
    \item \textsl{Word Embedding}: Word embedding is a technique that maps words into a high-dimensional vector space, where semantically similar words are positioned closer together in the vector space. This approach captures semantic relationships and contextual information between words. Common word embedding methods include Word2Vec \cite{b9}.
    \begin{enumerate}
        \item \textsl{Word2Vec}: Word2Vec is a technique that maps words into a high-dimensional vector space, capable of capturing semantic relationships and contextual information between words. This technique consists of two main models: CBOW (Continuous Bag of Words) and Skip-gram.
        \begin{itemize}
            \item The CBOW model trains word vectors by predicting a target word given its surrounding context. It uses the context words to predict the central word, enabling the model to effectively capture semantic relationships between words.
            \item The Skip-gram model operates in the opposite manner, training word vectors by predicting the context words given a target word. The Skip-gram model is better suited for handling small-scale datasets and excels at capturing semantic relationships between words.
        \end{itemize}
    \end{enumerate}
    
    In the study by DIANHUI MAO et al. \cite{b10}, the Word2Vec technique was applied to vectorize datasets of smart contracts. By training a Word2Vec model, the research converted key terms in smart contracts into high-dimensional vectors. These word vectors were then used for clustering analysis and automated code generation, improving the efficiency of smart contract design. The results demonstrated that by transforming keywords in smart contracts into high-dimensional vectors, the research team was able to capture the semantic features of the contracts and leverage these features for automated contract code generation.
    \item \textsl{TF-IDF}: TF-IDF(Term Frequency-Inverse Document Frequency) \cite{b11} is a commonly used statistical method for evaluating the importance of words within a collection of documents. It combines the term frequency (TF), which measures how often a word appears in a specific document, with the inverse document frequency (IDF), which reflects how rare the word is across all documents, to determine the word's significance.
    \begin{enumerate}
        \item \textsl{Term Frequency(TF)}: The TF measures the frequency of a word's occurrence within a document as shown in the equation below:
        \begin{align}
            tf_{t, d} = \frac{n_{t, d}}{\sum_{k=1}^{T}n_{k, d}}
        \end{align}
        \item \textsl{Inverse Document Frequency(IDF)}: IDF measures the importance of a word across the entire document collection as shown in the equation below:
        \begin{align}
            idf_{t} = log(\frac{D}{d_{t}})
        \end{align}
        \item \textsl{TF-IDF}: The TF-IDF value is obtained by multiplying TF and IDF as shown in the equation below:
        \begin{align}
            score_{t, d} = tf_{t, d} \times idf_{t}
        \end{align}
    \end{enumerate}
    In the study by DIANHUI MAO et al., TF-IDF was used to extract features from a dataset of smart contract source code retrieved from the Ethereum blockchain. The study utilized TF-IDF to calculate the keyword weights in the smart contracts and combined it with the K-means++ \cite{b12} clustering algorithm to classify and analyze the features of the smart contracts. This approach effectively identifies key features within the smart contracts and classifies them into different categories for further analysis and automated generation of corresponding contract source code.
\end{enumerate}

\section{Dataset and Data Processing}
In this work, we aim to develop a model capable of detecting vulnerabilities in real-world smart contracts. To achieve this, we decided to collect available smart contracts from the Ethereum public-chain and real-world financial vulnerability datasets from GitHub.

\subsection{Testing Dataset}

The testing data utilized in this study originates from the Ethereum public-chain. Historically, there have been two primary methods for collecting smart contract data. The first involves registering a cryptocurrency wallet and synchronizing a Full Node, which downloads the most recent transactions, typically spanning the last 128 blocks, to the user's device. The second method involves synchronizing an Archive Node, which downloads the entire transaction block data from 2015 to the present onto the user's device.

Both approaches involve downloading the complete transaction blocks from the Ethereum chain. However, for our analysis, the only information of real value is often the Solidity source code of smart contracts. Storing transaction data, ledger information, and other block-related details incurs significant storage costs, which appears inefficient; Therefore, we explored alternative methods to efficiently retrieve the Solidity source code of smart contracts. During this process, we discovered that public-chain websites not only provide access to transaction records but also offer APIs for developers to interact with smart contract data. As a result, we utilized these APIs to implement a Solidity source code crawler.

We developed a dynamic web crawler script to interact with platform APIs and retrieve smart contract code. In accordance with the API specifications, we registered accounts to request JSON data, extracted the Solidity source code, and stored they in our local MongoDB database. 

Simultaneously, we computed the hash value of each smart contract and compared it against entries in the smart contract hash table to exclude contracts that were already downloaded and stored in our database.

Certain well-known smart contract code may be directly copied by developers or companies onto public blockchains with lower transaction fees. For instance, developers or companies aiming to issue virtual tokens often adopt widely-used token contract standards, such as ERC-20. Consequently, data retrieved from different public-chains may contain identical contract code, though other fields, such as compiled version information, may differ. These discrepancies can impact hash computation results. To address this issue, we compute hashes exclusively based on the source code, thereby eliminating such inconsistencies.

Thus, we selected seven Ethereum public-chain platforms and collected over ten thousand Solidity source codes of smart contracts:

\begin{itemize}
    \item Etherscan
    \item BscScan
    \item PolygonScan
    \item CeloScan
    \item FTMScan
    \item Optimism
    \item Arbiscan
\end{itemize}

\subsection{Training Dataset}
During our investigation and literature review, we identified two datasets on the internet that are more suitable for unsupervised learning methods: the SmartBugs-curated dataset \cite{b13} and the SolidiFI-benchmark dataset \cite{b14}, so the training data utilized in this paper combine these two dataset  and develop the static analysis model for detection five vulnerabilities
\begin{itemize}
    \item Reentrancy
    \item Access Control
    \item Timestamp Dependency
    \item tx.origin
    \item Unchecked Low Level Call
\end{itemize}

\subsection{Data Processing}
To construct the clean training dataset related to the SmartBugs-curated and SolidiFI datasets, we selected 47,398 distinct Ethereum smart contracts from Etherscan, collected from the SmartBugs Wild and SolidiFI datasets. We then employed the static analysis tool Slither \cite{b15} and the dynamic analysis tool Mythril \cite{b16} to scan the source code, excluding contracts with questionable scan results in order to create a dataset as free from vulnerabilities as possible.

\begin{figure}[htbp]
\centerline{\includegraphics[width=0.7\columnwidth]{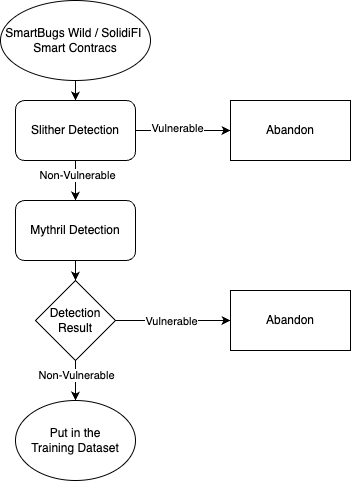}}
\caption{Vulnerability-Free Dataset Construction Flowchart.}
\label{data-processing}
\end{figure}

The dataset cleaning process was performed using Slither first, followed by Mythril, as this sequence accelerated the scanning procedure, as shown in Figure~\ref{data-processing}. Additionally, according to the research by Ka-Wing Man, clustering using a 30\% vulnerable data and 70\% non-vulnerable data ratio yields favorable results. Therefore, we adopted a 30/70 ratio to construct our training dataset.

\section{IMPLEMENTATION}
\begin{figure}[htbp]
\centerline{\includegraphics[width=0.7\columnwidth]{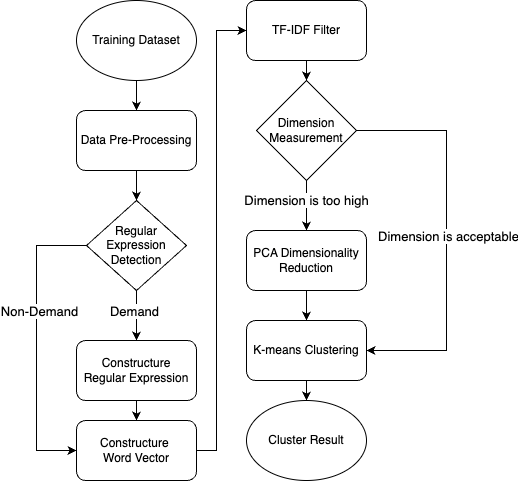}}
\caption{Unsupervised Learning Analysis System Flowchart}
\label{k-means}
\end{figure}

In this work, we develop an unsupervised learning approach using K-means clustering to classify smart contracts into different clusters. This method determines whether a contract lies near the normal center or is an outlier, indicative of an abnormal smart contract. The flowchart, shown in Figure~\ref{k-means}, outlines the six steps involved in detecting vulnerabilities.

Our approach consists of six methods:
\begin{enumerate}
    \item Data Pre-Processing
    \item Regular Expression Detection
    \item Word Vector Construction
    \item TF-IDF Filter
    \item Principal Component Analysis (PCA) Dimensionality Reduction
    \item K-means Clustering
\end{enumerate}

\subsection{Data Pre-Processing}
The source code of the smart contract must undergo preprocessing. The purpose of this step is to remove noise from the contracts, including comments, fixed keywords, symbols, and unnecessary whitespace, to provide a standardized and easily processable data format for subsequent steps.

\begin{table}[htbp]
\caption{Solidity Keywords}
\begin{center}
\begin{tabular}{|c|c|c|c|}
\hline
`pragma` & `import` & `contract` & `interface` \\
\hline
`library` & `struct` & `enum` & `function`  \\
\hline
`event` & `error` & `using` & `for` \\
\hline
`constructor` & `mapping` & `address` & `bool` \\
\hline
`string` & `var` & `bytes` & `uint` \\
\hline
`int` & `if` & `else` & `while`\\
\hline
`do` & `break` & `continue` & `return` \\
\hline
`throw` & `emit` & `public` & `private` \\
\hline
`internal` & `external` & `constant` & `immutable` \\
\hline
`view` & `pure` & `virtual` & `override` \\
\hline
`storage` & `memory` & `calldata` & `try` \\
\hline
`catch` & `revert` & `assert` & `require` \\
\hline
`new` & `delete` & `this` & `solidity`\\
\hline
\end{tabular}
\label{solidity-keywords}
\end{center}
\end{table}

We convert the smart contract .sol files stored as .txt text files, where each line in the text file represents a single keyword, then remove the fixed, commonly used keywords in Solidity since these words are reserved keywords in the programming language as Table~\ref{solidity-keywords}, their removal makes it easier to highlight the unique keywords of each contract.

\begin{algorithm}
\caption{Remove non-demand noise}\label{alg:noise-remove}
\begin{algorithmic}[1] 
    \Require Solidity source code.
    \Ensure Code no comments, punctuation, redundant space.
    \State $regex$ $\leftarrow$ r`//.*?\textbackslash n$\vert$/\textbackslash*.*?\textbackslash*/',"
    \While{$punctuation$}
        \If{$comments$}
            \State $code$ $\leftarrow$ re.sub($regex$, $code$, flags = $code$.DOTALL)
        \Else
            \State $code$ $\leftarrow$ $code$.replace(`//`, ` `).replace(`/*`,` `).replace(`*/`,` `)
        \EndIf
    \State $no\_punctuation$ $\leftarrow$ $code$.translate(translator)
    \State $no\_slash$ $\leftarrow$ $no\_punctuation$.replace(`/`,` `)
    \State $stripped\_code$ $\leftarrow$ $no\_slash$.strip().replace(`\textbackslash n`, ` `)
    \State $single\_space$ $\leftarrow$ re.sub(`+`, ` `, $stripped\_code$)
    \EndWhile
\State \textbf{Return} $stripped\_code$
\end{algorithmic}
\end{algorithm}

Also we should consider the comments, operators, and unnecessary whitespace need removed using Algorithm~\ref{alg:noise-remove}.

\subsection{Regular Expression Detection}
Given that our approach predominantly relies on extracting keywords from contracts as features for clustering to identify vulnerabilities, it is limited in its ability to detect certain types of vulnerabilities, such as reentrancy, solely through keyword analysis. To address this limitation, we integrate a regular expression-based detection mechanism for vulnerabilities including Reentrancy, Timestamp Dependency, tx.origin usage, and Unchecked Low-Level Calls. Contracts that match predefined regular expression patterns are flagged, and corresponding word vectors for the identified keywords are directly incorporated into the training set, bypassing the traditional TF-IDF-based feature selection process. 

\begin{algorithm}
\caption{Regular expression pattern detection}\label{alg:regular-expression}
\begin{algorithmic}[1] 
    \Require Solidity source code.
    \Ensure The codes match the regular expression.
    \State pattern $\leftarrow$ re.compile(Match the patterns associated with each vulnerability)
\end{algorithmic}
\end{algorithm}

Given the highly variable nature of smart contract source code, constructing a universal regular expression pattern is infeasible. As a result, we perform scanning on the preprocessed text files. We then construct an array matching the size of the dataset to read the previously preprocessed files. The files are scanned using regular expressions, and if a specific pattern is matched, the corresponding element in the array is set to 1.

\subsubsection{Reentrancy Patterns}

\begin{itemize}
    \item Trigger: r`\textbackslash b(call)\textbackslash b`, re.IGNORECASE)
    \item Balance: r’\textbackslash b(balance$\vert$balances)\textbackslash b’, re.IGNORECASE)
\end{itemize}

\subsubsection{Timestamp  Dependency Pattern}
r``((\textbackslash bnow\textbackslash b)$\vert$(\textbackslash bblock\textbackslash.timestamp\textbackslash b))``)

\subsubsection{tx.origin Pattern}

r``\textasciicircum\textbackslash{}s*(require$\vert$if)\textbackslash s*\textbackslash (\textbackslash s*tx\textbackslash.origin\textbackslash s \\ *[!=]=?\textbackslash s*[\textbackslash w\textbackslash d\_]+\textbackslash s*\textbackslash)``
, re.MULTILINE)

\subsubsection{Unchecked Low Level Call}
\begin{itemize}
    \item Prefixes Pattern: r`\textbackslash b(require$\vert$if$\vert$bool$\vert$success)\textbackslash b`
    \item Postfixes Pattern: r`\textbackslash.(call$\vert$value$\vert$callcode$\vert$delegatecall$\vert$staticcall\\$\vert$send)\textbackslash(`
\end{itemize}

The condition for a reentrancy vulnerability is that the contract's state is changed only after a transfer is executed, with state changes predominantly associated with the balance(s) keyword. The choice of call as the transfer keyword is due to the fact that call does not impose a limit on the gas fee, which increases the likelihood of a successful reentrancy attack. Therefore, our pattern is designed to detect the call keyword, and if the balance(s) keyword appears within five lines, the corresponding element in the array for that contract is flagged.

\begin{algorithm}
\caption{Detection the Reentrancy pattern}\label{alg:detection-reentrancy}
\begin{algorithmic}[1] 
    \Require Solidity source code
    \Ensure The codes match the reentrancy regular expression
    \For{$line$ $\in$ code}
        \If{$line\_buffer$}
            $line\_buffer$.append($line$)
            \If{$balance\_patern$.search($\ell$) $\forall \ell \in line\_buffer$}
                \State \textbf{Return $1$}
            \ElsIf{len($line\_buffer \geq 5$)}
                \State $line\_buffer$.pop(0)
            \EndIf
        \EndIf
        \If{$trigger\_pattern$.search($line$)}
            $line\_buffer \leftarrow [line]$ 
        \EndIf
    \EndFor
\State \textbf{Return} $line\_buffer$
\end{algorithmic}
\end{algorithm}

\subsection{Word Vector Construction}
In our approach, we detect vulnerable sentences from Solidity source code and transform them into data points. To represent the words, we construct word vectors using Gensim's Word2Vec \cite{b17}. Several parameters need to be considered:
\begin{itemize}
    \item \textsl{documents}: Corpus to be trained.
    \item \textsl{vector\_size}: The size of the word vector: larger dimensions can capture more semantic relationships but require more computational resources.
    \item \textsl{window}: Window size, which represents the maximum distance between the current word and the predicted word.
    \item \textsl{min\_count}: Ignore all words with a frequency lower than this value.
    \item \textsl{workers}: The number of worker threads for training. Gensim’s Word2Vec can utilize multiple cores to accelerate training.
    \item \textsl{sg}: Define the training algorithm. 0 represents CBOW (Continuous Bag of Words), and 1 represents Skip-gram. Skip-gram typically performs better on smaller datasets, while CBOW is faster on larger datasets.
    \item \textsl{epochs}: The number of iterations during the training process.
\end{itemize}
We only need to construct the vector for each word, rather than making predictions, so adjustments to the detailed parameters will not significantly impact the subsequent experiments.

\begin{algorithm}
\caption{TF-IDF}\label{alg:tf-idf}
\begin{algorithmic}[1]
    \Require Documents
    \Ensure The TF-IDF model through a specific corpus
    \State $dictionary \leftarrow$ Dictionary($documents$)
    \State $corpus \leftarrow$ [$dictionary$.doc2bow($text$) $\forall text \in documents$]
    \State $tfidf\_model \leftarrow$ TfidfModel($corpus$)
    \State \textbf{Return $tfidf\_model$}
\end{algorithmic}
\end{algorithm}

\subsection{TF-IDF Filter}
After constructing the word vectors, we use Gensim's TF-IDF \cite{b18} for feature selection to determine which keyword vectors will be used to represent the features of the contract for the subsequent clustering step, so construct the corpus for TF-IDF calculation, and then train the TF-IDF model.

\begin{algorithm}
\caption{Filter the keyword and construct the word list}\label{alg:filter}
\begin{algorithmic}[1]
    \Require The Corpus
    \Ensure The word vectors via TF-IDF
    \State $word\_vector\_tfidf \leftarrow \{\varnothing\}$
    \State $counter \leftarrow 0$
    \For{$doc \in corpus$}
        \For{$word\_id, tfidf\_score \in tfidf\_model[doc]$}
            \If{$tfidf\_score > threshold$}
                \State $word \leftarrow dictionary[word\_id]$
                \If{$word \in$ word2vec\_model.wv}
                    $word\_vector\_tfidf[word] \leftarrow$ word2vec\_model.wv[$word$]
                \EndIf
            \EndIf
        \If{$regex\_result[counter]$}
            \If{$dictionary[word\_id] \in$ `$call$`}
                \State $word \leftarrow dictionary[word\_id]$
                \If{$word \in$ word2vec\_model.wv}
                    \State $word\_vector\_tfidf[word] \leftarrow$ word2vec\_model.wv[$word$]
                \EndIf
            \EndIf
        \EndIf
        \EndFor
        \State $counter++$
    \EndFor
\end{algorithmic}
\end{algorithm}

We use the tfidf\_score as a threshold, where word vectors with a score above this value are added to the dataset for clustering. A higher TF-IDF weight indicates a higher uniqueness of the keyword, making it a better representation of the contract's features. However, setting the threshold too high may result in insufficient features, while setting it too low may lead to an excessive number of features, reducing the effectiveness of clustering. Therefore, the tfidf\_score should be adjusted according to the experimental data. Word vectors that meet the threshold are added to the word\_vector\_tfidf list for the next phase of clustering. When reading the vectors of each contract in the corpus, we also check the results of the regular expression detection for that contract. If the array element is 1, the keyword's word vector is directly added to the word\_vector\_tfidf list.

Since a single contract may encounter issues with duplicate keywords or may not contain keywords that meet the TF-IDF threshold, we address these two problems as follows: first, duplicate keywords are removed, retaining only one word vector; for contracts without any keywords meeting the threshold, a zero vector is used as padding. The processed vectors are appended to the document\_vectors list. Finally, the trained Word2Vec model and the feature list for each contract are saved to a file for subsequent testing and clustering, eliminating the need for retraining.

\begin{algorithm}
\caption{Remove the redundant keyword and padding}\label{alg:remove-redundant}
\begin{algorithmic}[1]
    \Require Documents
    \Ensure The document vectors after remove redundant keywords
    \State $document\_vectors \leftarrow [\varnothing]$
    \For{$doc \in documents$}
        \State $unique\_words \leftarrow$ set($doc$)
        \State $filtered\_vector \leftarrow [(word, word\_vector\_tfidf[word]) \forall word \in unique\_word$ \textbf{if} $word \in word\_vector\_tfidf]$
        \If{$filtered\_vectors$}
            \State $vectors\_only \leftarrow [vec \forall \_, vec \in filtered\_vecotrs]$
            \State $doc\_vector \leftarrow$ np.mean($vectors\_only, axis \leftarrow 0)$
            \State $document\_vectors$.append($doc\_vector$)
        \Else
            \State $zeor\_vector \leftarrow$ np.zeros(word2vec\_model.vector\_size)
            \State $document\_vectors$.append($zero\_vector$)
        \EndIf
    \EndFor
\State \textbf{Return} $document\_vectors$
\end{algorithmic}    
\end{algorithm}

\subsection{Principal Component Analysis (PCA) Dimensionality Reduction}

Before clustering, if the dimensionality of the word vectors is determined to be too high, PCA is applied for dimensionality reduction to improve the performance of the subsequent k-means clustering.

\begin{algorithm}
\caption{PCA Dimensionality Reduction}\label{alg:PCA}
\begin{algorithmic}[1]
    \Require Word Vectors
    \Ensure The reduced dimension of word vectors
    \State $X\_mean \leftarrow$ torch.mean($X, dim \leftarrow 0$)
    \State $X\_centered \leftarrow X - X\_mean$
    \State $U, S, V \leftarrow$ torch.svd($X\_centered$)
    \State \textbf{Return} torch.matmul($X\_centered, V[:, :$num\_components])
\end{algorithmic}
\end{algorithm}

\subsection{K-means Clustering}
We use the PyTorch k-means library for clustering. According to the research by Ka-Wing Man, combining k-means with PCA for dimensionality reduction can achieve excellent performance, so in this work we develop this approach. 

Since the initial cluster centers in k-means are randomly selected during each execution, the results may vary across runs. To ensure reproducibility in our experiments, we fix the PyTorch seed, ensuring consistent selection of cluster centers in each execution.

Then, the previously saved document\_vectors training dataset file is loaded, and a PyTorch tensor is constructed for k-means clustering, some important parameter are:
\begin{itemize}
    \item \textsl{num\_clusters}: The number of clusters. This parameter needs to be tested and adjusted based on the specific application scenario and data characteristics to achieve optimal results.
    \item \textsl{max\_iterations}: The maximum number of iterations for the k-means algorithm, set to prevent prolonged execution without convergence in certain scenarios.
    \item \textsl{document\_vectors\_tensor}: The dataset to be clustered, which in this case consists of the previously processed word vectors.
    \item \textsl{distance}: The method for measuring distance, in this case, Euclidean Distance, which is a commonly used metric in k-means clustering. The Euclidean Distance calculation formula is as follows:
    \begin{align}
        d(p, q) = \sqrt{\sum_{i}(p_{i} - q_{i})^{2}}
    \end{align}
    \item \textsl{device}: The device used to run k-means. First, it checks for the availability of a CUDA-enabled GPU; if available, the GPU is used; otherwise, the CPU is utilized.
    
    \begin{algorithm}
    \caption{K-means Clustering}\label{alg:k-means}
    \begin{algorithmic}[1]
        \Require Document vectors tensor
        \Ensure Clustering the data
        \State $num\_clusters \leftarrow$ Heuristically choose the number of clusters
        \State $max\_iterations \leftarrow$ Heuristically choose the maximum number of iterations
        \State $cluster\_ids\_x, cluster\_centers \leftarrow$ kmeas($X \leftarrow document\_vectors\_tensor, num\_clusters \leftarrow num\_clusters, distance \leftarrow$ `euclidean`, $device \leftarrow$ torch.device(`cuda` \textbf{if} torch.cuda.is\_available() \textbf{else} `cpu`), $iter\_limit \leftarrow max\_iterations$)
    \end{algorithmic}
    \end{algorithm}
\end{itemize}

After clustering is completed, the results are statistically analyzed and output, cluster\_ids\_x represents the cluster number assigned to each document. Since the documents are constructed in sequence, the first 30\% contain vulnerable phrases, while the remaining 70\% are non-vulnerable. Therefore, the clustering results can be statistically analyzed based on the sequence of cluster numbers.

All the source codes and results of this paper are available at https://github.com/hshuang99/EthCluster

\section{EXPERIMENTAL RESULTS}
In this section, we present the experimental results for vulnerabilities such as Reentrancy, Access Control, Timestamp Dependency, tx.origin, and Unchecked Low-Level Call. These vulnerabilities are classified using our Unsupervised Learning Static Analysis System. The effectiveness of the unsupervised learning approach depends on the characteristics of the dataset. Therefore, we have heuristically determined the appropriate cluster size for each    vulnerability.

\subsection{Evaluation Metrics}
When an anomaly detection algorithm is applied, four possible cases must be considered. The first case is the correct detection of vulnerabilities, where the detected anomalies in the data correspond exactly to abnormalities in the process. This is referred to as a True Positive (TP). The second case is the correct identification of normal behavior, where no abnormalities are detected, referred to as a True Negative (TN).

The third case is a False Positive (FP), where the process remains normal, but unexpected data values are incorrectly identified as anomalies. Lastly, the False Negative (FN) occurs when the process becomes abnormal, but the resulting anomalies are not registered in the data.

These classification results are utilized to construct the confusion matrix. Consequently, our experiment employs widely accepted evaluation criteria for vulnerability detection tasks, using assessment indices such as:

\begin{itemize}
    \item \textsl{Accuracy (ACC)}: Evaluates the precise classification results across all samples.
    \begin{align}
        Accuracy = \frac{TP+TN}{TP+TN+FP+FN}
    \end{align}
    However, it is important to note that accuracy in anomaly detection may provide a misleading prediction when there is a significant imbalance between the number of positive and negative examples.
    \item \textsl{Precision(P)}: Evaluates the accuracy of detecting vulnerable samples.
    \begin{align}
        Precision = \frac{TP}{TP+FP}
    \end{align}
    \item \textsl{Recall (R)}: Measures the accuracy of predicting vulnerable samples.
    \begin{align}
        Recall = \frac{TP}{TP+FN}
    \end{align}
    \item \textsl{F-measure (F)}: Evaluates both precision and recall to optimize the detection results.
    \begin{align}
        F-measure = \frac{2TP}{2TP+FN+FP}
    \end{align}
\end{itemize}

\subsection{Evaluation Results}
In our prediction model, we utilized 1,000 real-world smart contracts to test our clustering approach. The heuristic selection of the randomness parameter for the k-means clustering algorithm was set to 1,194. Table~\ref{tab:clustering-results} presents the clustering parameters for each vulnerability.

\begin{table}[htbp]
\caption{Clustering Parameters}
\begin{center}
\begin{tabular}{|c|c|c|c|}
\hline
\textbf{Vulnerability} & \textbf{Vector Size} & \textbf{TF-IDF Score} & \textbf{Cluster}\\
\hline
Reentrancy & 10 & 0.7 & 5 \\
\hline
Access Control & 50 & 0.3 & 3 \\
\hline
Timestamp Dependency & 300 & 0.7 & 6\\
\hline
tx.origin & 300 & 0.7 & 6\\
\hline
Uncheck Low Level Call & 100 & 0.7 & 8\\
\hline
\end{tabular}
\label{tab:clustering-results}
\end{center}
\end{table}

Subsequently, we present the confusion matrix for each vulnerability in Table~\ref{tab:confusion-matrix}, with the values derived from the SmartBugs-curated dataset and the SolidiFI benchmark dataset.

\begin{table}[htbp]
\scriptsize
\caption{Confusion Matrix}
\begin{center}
\begin{tabular}{|c|c|c|c|c|}
\hline
\textbf{Vulnerability} & \textbf{TP} & \textbf{FP} & \textbf{FN} & \textbf{TN}\\
\hline
Reentrancy & 81 (30\%) & 7 (3\%) & 0 (0\%) & 182 (67\%) \\
\hline
Access Control & 18 (30\%) & 0 (0\%) & 8 (13.3\%) & 34 (56.67\%) \\
\hline
Timestamp Dependency & 49 (27\%) & 2 (1\%) & 6 (3.3\%) & 127 (70\%)\\
\hline
tx.origin & 50 (30\%) & 0 (0\%) & 0 (0\%) & 117 (70\%)\\
\hline
Uncheck Low Level Call & 52 (29.9\%) & 8 (4.6\%) & 0 (0\%) & 114 (65.5\%)\\
\hline
\end{tabular}
\label{tab:confusion-matrix}
\end{center}
\end{table}

Furthermore, our detection results for each vulnerability are presented in Table~\ref{tab:detection-results}.

\begin{table}[htbp]
\caption{The detection results of vulnerability}
\begin{center}
\begin{tabular}{|c|c|c|c|c|}
\hline
\textbf{Vulnerability} & \textbf{ACC(\%)} & \textbf{P(\%)} & \textbf{R(\%)} & \textbf{F(\%)}\\
\hline
Reentrancy & 97.41 & 92.05 & 100 & 95.86 \\
\hline
Access Control & 86.67 & 100 & 69.23 & 81.82 \\
\hline
Timestamp Dependency & 95.65 & 96.08 & 89.09 & 92.45\\
\hline
tx.origin & 100 & 100 & 100 & 100\\
\hline
Uncheck Low Level Call & 95.4 & 86.67 & 100 & 92.86\\
\hline
\end{tabular}
\label{tab:detection-results}
\end{center}
\end{table}

We generate the result figures using Matplotlib, as illustrated below:

\begin{figure}[htbp]
\centerline{\includegraphics[width=0.7\columnwidth]{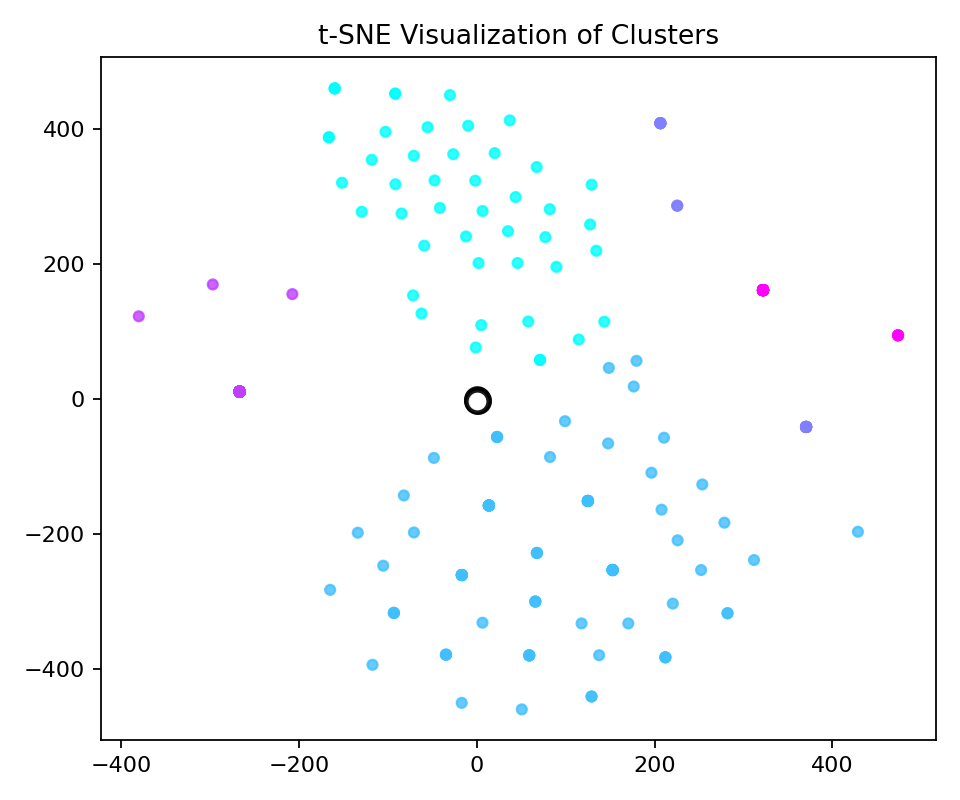}}
\caption{The t-SNE figure of Reentrancy.}
\label{fig:reentrancy-tsne}
\end{figure}

\begin{figure}[htbp]
\centerline{\includegraphics[width=0.7\columnwidth]{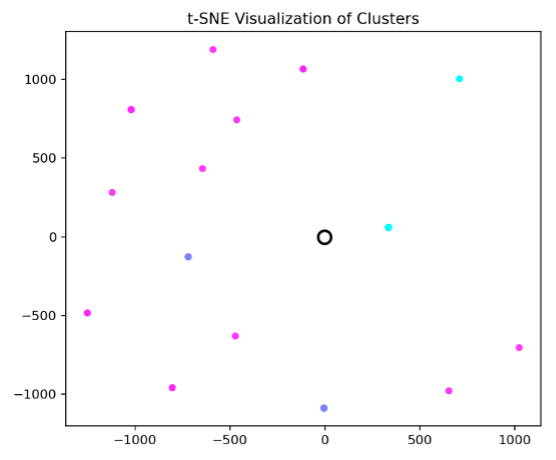}}
\caption{The t-SNE figure of Access Control.}
\label{fig:access-control-tsne}
\end{figure}

\begin{figure}[htbp]
\centerline{\includegraphics[width=0.7\columnwidth]{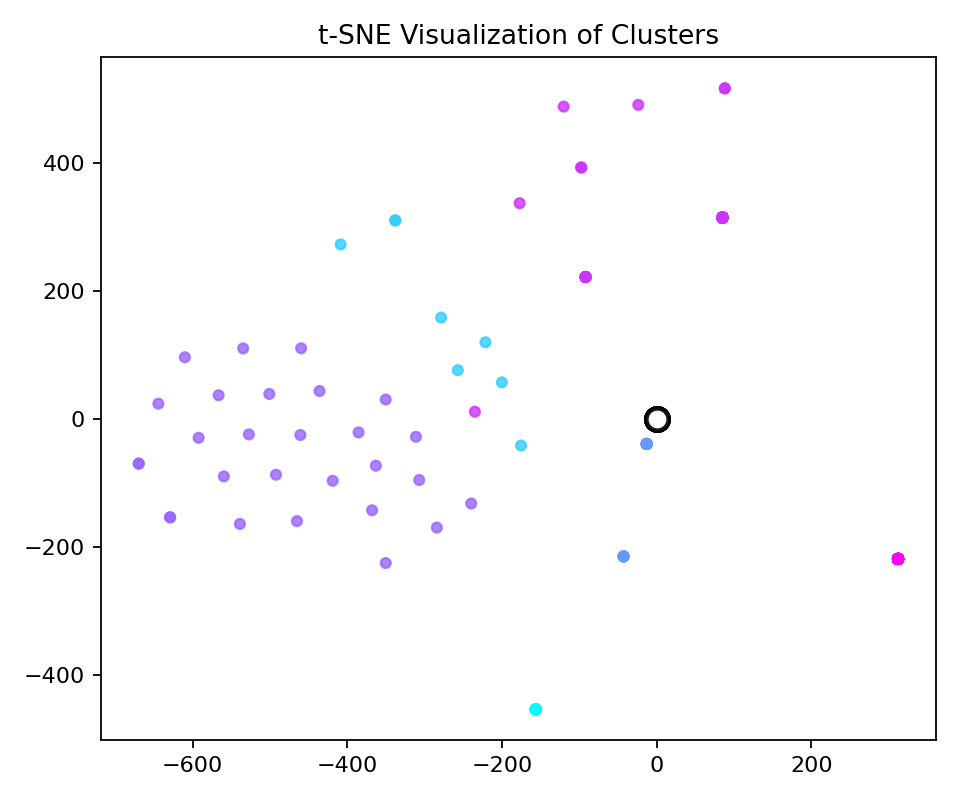}}
\caption{The t-SNE figure of Timestamp Dependency.}
\label{fig:timestamp-dependency-tsne}
\end{figure}

\begin{figure}[htbp]
\centerline{\includegraphics[width=0.7\columnwidth]{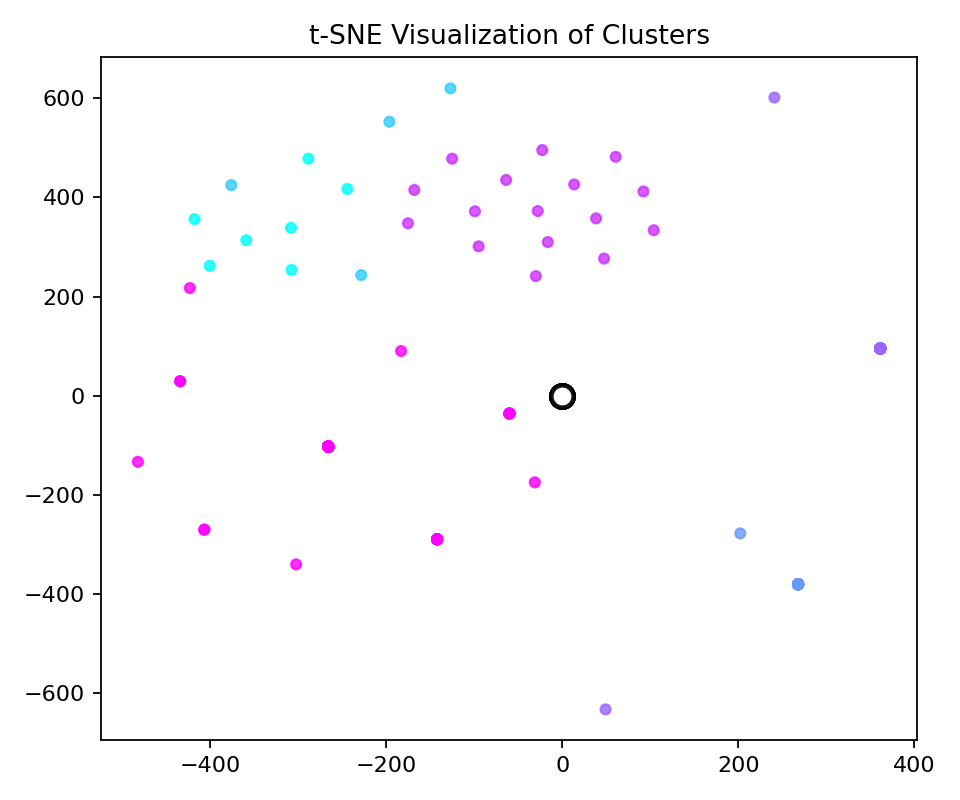}}
\caption{The t-SNE figure of tx.origin.}
\label{fig:tx-origin-tsne}
\end{figure}

\begin{figure}[htbp]
\centerline{\includegraphics[width=0.7\columnwidth]{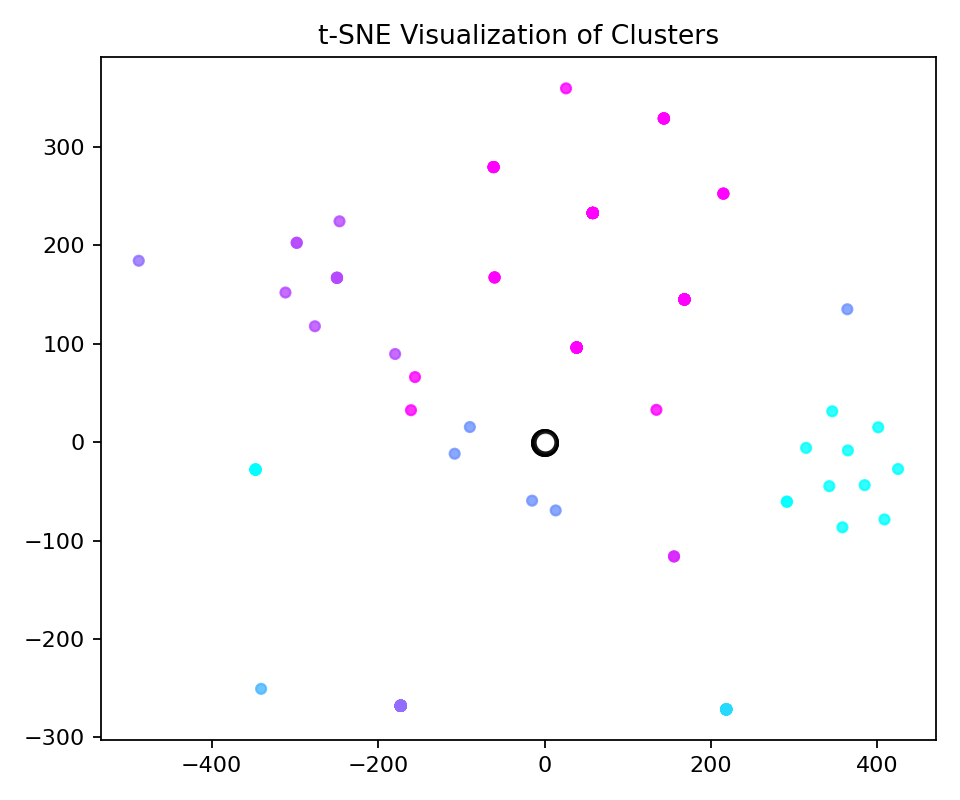}}
\caption{The t-SNE figure of Uncheck Low Level Call.}
\label{fig:un-checked-low-level-calls-tsne}
\end{figure}

\section{DISCUSSION}
Data has always been the most critical resource for both machine learning and deep learning methods, as sufficient data is essential for precise training. In this year’s project, we began by collecting data to create a model that better reflects real-world scenarios. To achieve this, we utilized dynamic web crawlers on Ethereum public chain websites to extract Solidity source code from smart contracts deployed by various companies as testing data. The speed of the crawler is determined by machine performance and network bandwidth. Although this approach does not require downloading entire blocks through a wallet connected to the Ethereum chain as in previous methods, new smart contracts are deployed weekly. As a result, the number of contracts stored in MongoDB continues to grow, requiring increasingly larger storage capacities.

Moreover, as the number of contracts increases, many well-known token contract standards, such as ERC-20, are often duplicated with only minor changes, such as the issuing company’s name. This necessitates frequent comparisons of contract content within the database to identify and exclude duplicates. Failure to do so could compromise the model’s testing data and lead to issues such as overfitting.

For training data, real-world samples of actual incidents are required. However, such datasets are exceedingly rare. Our investigation identified only two public datasets online, which together cover only ten types of vulnerabilities and contain merely a few hundred contracts available for training. We also attempted to use generative AI models, including GPT-4, GPT-4-turbo, GPT-4o, and Gemini Pro, to generate vulnerable contract samples. Despite repeated testing and some improvements in generating smart contracts with each  model update, the uniqueness of each generated contract remains insufficient. This lack of diversity can lead to over-optimization during training and negatively impact
subsequent testing. Additionally, generative AI is limited by constraints on length and speed, making it impractical for generating large volumes of training data.

In contrast, our weekly updates of testing data collected via crawlers have already exceeded over 100,000 samples. This imbalance in the ratio of training to testing data is a common issue in deep learning, particularly in anomaly detection methods. As previously mentioned, the accuracy of model predictions heavily depends on the quality of the dataset. Therefore, addressing the challenges of data imbalance in vulnerability detection will require the continuous development of new methods, both now and in the future.

\section{CONCLUSION}
In this paper, we decided to adopt an unsupervised deep learning approach using clustering algorithms to establish a static analysis system. This decision was made due to the significant disparity between the number of vulnerable contract samples and the available Ethereum smart contract test data. We utilized dynamic web crawlers to send API requests to target websites, retrieving smart contract source code from the Ethereum public blockchain. The retrieved data was stored in MongoDB for subsequent testing purposes.

For training vulnerable contract samples, we identified two publicly available datasets: Smartbugs-curated and SolidiFI-benchmark. We also developed a model based on clustering algorithms to distribute the contract data within the datasets across various clusters. In this model, if the cluster center represents normal data, the surrounding data points are also considered normal and classified as non-vulnerable contracts. Conversely, if a contract does not cluster around normal data points and is identified as an outlier, it is classified as abnormal and considered a vulnerable contract.

In the domain of anomaly detection, a variety of methods have been investigated, including K-Nearest Neighbor, Autoencoder, DRAEM, and GCAD, among others. To mitigate the challenges posed by imbalanced datasets, recent research has introduced approaches such as Few-Shot Learning, Zero-Shot Learning, and Contrastive Language-Image Pre-training (CLIP), all of which have shown
significant promise in enhancing efficiency. These methodologies present potential avenues for future advancements and optimization of the system.

\section*{Acknowledgment}
This research was supported in part by the Ministry of Science and Technology, Taiwan, under Project MOST 111-2221-E-007-078-MY3.

\end{document}